\documentclass[review]{elsarticle}

\usepackage{amsmath}

\journal{Wave Motion}

\begin{document}

\begin{frontmatter}

\title{Random matrix theory for an adiabatically-varying oceanic acoustic waveguide}

\author{D.V. Makarov}
\address{Laboratory of Nonlinear Dynamical Systems, POI FEB RAS, 43 Baltiyskaya Str., 690041, Vladivostok, Russia}




\begin{abstract}

Problem of sound propagation in the ocean is considered.
A novel approach of K. Hegewisch and S. Tomsovic for statistical modelling of acoustic wavefields in the random ocean is examined.
The approach is based on construction of a wavefield propagator by means of random matrix theory.
It is shown that this approach can be generalized onto acoustic waveguides with adiabatic longitudinal variations.
Efficient generalization is obtained by means of stepwise approximation of the propagator.
Accuracy of the generalized approach is confirmed numerically for a model of an underwater sound channel crossing a cold synoptic eddy.
It is found that the eddy leads to substantial suppression of sound scattering. 
\end{abstract}

\begin{keyword}
 ocean acoustics \sep  sound scattering \sep random matrix theory \sep wavefield propagator \sep normal modes 
 \MSC[2010] 74J20\sep  65C05
\end{keyword}

\end{frontmatter}


\section{Introduction}

For many decades now wave scattering in random media is one of the most important problems of wave theory.
From the viewpoint of practical applications, it is thought of as an adverse process worsening signal-to-noise ratio.
In the context of long-range sound propagation in the ocean, volume scattering on random sound-speed inhomogeneity
severely delimits possibilities of hydroacoustical tomography \cite{TT96}. 
Such inhomogeneity is commonly caused by oceanic internal waves. Internal-wave-induced sound-speed variations are 
usually small, inferring only forward scattering,
but their accumulated long-range effect can be very substantial, as it is confirmed by experiments \cite{SLICE89,AET,AST}. 

In the ray-based description, internal waves give rise to Lyapunov instability and chaos of sound rays \cite{Review03,RayWave,UFN}.
The phenomenon of ray chaos is mathematically equivalent to dynamical chaos in classical physics.
Following this analogy, wavefield manifestations of ray chaos, commonly referred to as wave chaos, can be considered from the viewpont
of a more general paradigm of quantum chaos \cite{Stockman}.
This circumstance enables usage of well-developed methods of quantum chaos to the problem of long-range sound propagation.
In particular, we can mention phase space analysis using the Wigner function or its smoothed versions \cite{Sundaram_Zas,Okomelkova,Viro00,Viro05,Ac17},
Lyapunov analysis \cite{WT01}, entropy calculation \cite{MorCol}, periodic orbit theory \cite{Viro-scar,PRE76,SFU10}, theory of nonlinear resonance \cite{Viro01,Chaos,BV04}, 
and action-angle formalism \cite{Udo08}, to name a few.
One of the most novel approaches is based on the unitary propagator governing wave evolution 
within the narrow-angle approximation \cite{UFN,Levelspacing,Tomc11,PRE87,Hege13}.
Particularly, Hegewisch and Tomsovic have shown that such propagator can be constructed 
using random matrix theory (RMT), avoiding direct solution of the parabolic wave
equation \cite{Tomc11,Hege13}. 
Random matrices are utilized to describe mode coupling induced by scattering on the random inhomogeneity.
Hereafter we shall refer to this method as the Hegewisch-Tomsovic method.
Validity of the Hegewisch-Tomsovic approach was examined in \cite{JCA,UZMU-RMT}. It was shown that the random matrix modelling ensures
sufficient accuracy for signal frequencies of 50-100 Hz that are relevant for long-range propagation.

In the Hegewisch-Tomsovic method, solution of a wave equation is replaced
by multiplication of matrices. The matrix size is determined by number of propagating modes, therefore,
this method is extremely fast for low frequencies, if the background sound-speed profile doesn't depend on range.
 However, the latter condition is basically not satisfied in realistic oceanic environments.
 Ocean almost always has large-scale horizontal inhomogeneity due to temperature and bathymetric variations, presence of eddies and currents, Rossby waves,  e.t.c.
 The corresponding variations of a sound-speed profile are commonly very significant and cannot be considered as a small perturbation.
 Thus, applicability of the Hegewisch-Tomsovic method in natural experiments requires generalization
 onto waveguides with strong but adiabatic longitudinal variability.
 Unfortunately, an attempt to incorporate large-scale inhomogeneity directly to the original scheme of the method results in substantial growth
 of auxiliary computations. 
 In this way, the Hegewisch-Tomsovic method loses its important advantage, namely its speed.
Therefore one needs an optimized version of this method incorporating the effect of large-scale longitudinal variability.
The present work offers a pretty simple and robust way to resolve this problem. 

The paper is organized as follows. The next section contains brief description of the Hegewisch-Tomsovic method in the absence of adiabatic inhomogeneities.
Section \ref{Model} is devoted to the acoustic model used for numerical simulation.
Modification of the Hegewisch-Tomsovic method for waveguides with adiabatic inhomogeneity is presented in Section \ref{Adiabatic}.
In Section \ref{Numer}, we numerically examine validity of the modified Hegewisch-Tomsovic method by means of numerical simulation.
Section Discussion outlines some prospects for future research in this field.
In Conclusions, an account of the main results is presented.

\section{Hegewisch-Tomsovic method in the absence of large-scale sound-speed inhomogeneity}
\label{Absence}


Long-range wave propagation can be fairly modeled by means of the standard parabolic equation that takes into account only forward propagation.
Assuming cylindrical symmetry and neglecting azimuthal coupling, we can reduce the original three-dimensional problem to the two-dimensional one.
Then the parabolic equation can be written  in the following way:
\begin{equation}
\frac{i}{k_0}\frac{\partial\Psi}{\partial r}=
-\frac{1}{2k_0^2}\frac{\partial^2\Psi}{\partial
z^2}+\frac{n^2-1}{2}\Psi, 
\label{parabolic}
\end{equation}
where 
\begin{equation}
 k_0=\frac{2\pi f}{c_0},
 \label{k0}
\end{equation}
$z$ is ocean depth, $r$ is range,
$f$ is signal frequency, $c_0$ is a reference sound speed, and $n = n(r,z) = c_0/c(r,z)$ is refractive index.
In the small-angle approximation we have
\begin{equation}
 \frac{n^2(r,z)-1}{2} \simeq U(z) + V_{\text{lsc}}(r,z) + V_{\text{iw}}(r,z),
 \end{equation}
where 
\begin{equation}
U(z)=\frac{\Delta c(z)}{c_0},\quad V_{\text{lsc}}(r,z) = \frac{\delta c_{\text{lsc}}(r,z)}{c_0},\quad
V_{\text{iw}}(r,\,z)=\frac{\delta c_{\text{iw}}(r,\,z)}{c_0}.
\label{pot}
 \end{equation}
Here $\Delta c(z)$ is linked to the range-independent unperturbed sound-speed profile as $\Delta c(z) = c_{\text{unpert}}(z)-c_0$,
$\delta c_{\text{lsc}}(r,z)$ describes large-scale sound-speed inhomogeneity,
and $\delta c_{\text{iw}}(r,z)$ is a random sound-speed perturbation caused by internal waves.

Acoustic wavefield can be represented as sum over normal modes of the unperturbed waveguide
\begin{equation}
 \Psi(r,z) = \sum\limits_{m} a_m(r)\psi_m(z).
\end{equation}
The normal modes and the corresponding eigenvalues satisfy the Sturm-Liouville problem
\begin{equation}
-\frac{1}{2k_0^2}\frac{\partial^2\psi_m(z)}{\partial
 z^2}+U(z)\psi_m(z)=E_m\psi_m(z).
\label{StL}
\end{equation}
Solution of the parabolic equation (\ref{parabolic}) at the range $r=r_{\mathrm{f}}$
can be formally written in terms of an unitary propagator $\hat G$ acting as
\begin{equation}
 \Psi(r_{\mathrm{f}},z) = \hat G(r_0,r_{\mathrm{f}})\Psi(r_0,z).
 \label{evolution}
\end{equation}
Using the basis of normal modes, we can express the propagator $\hat G$ as a matrix $\mathbf{G}$ with elements
\begin{equation}
 G_{mn}(0,r_{\mathrm{f}})=\int \psi_m^*\hat G(0,r_{\mathrm{f}})\psi_n\,dz,
 \label{Gmn}
\end{equation}
where $\hat G(0,r_{\mathrm{f}})\psi_n$ is a solution of the parabolic equation at the range $r=r_{\mathrm{f}}$
for the initial condition $\Psi(r=0)=\psi_n$.
As long as the parabolic equation involves a random perturbation $V_{\text{iw}}(r,z)$,
the propagator matrix $\mathbf{G}$ is random as well.

For the sake of simplicity, we use idealistic perfectly-reflecting boundary conditions of the form
\begin{equation}
\left.\Psi\right\vert_{z=0} = 0,\quad
\left.\frac{d\Psi}{dz}\right\vert_{z=h}=0,
\label{BCs}
\end{equation}
where $h$ is depth of the ocean bottom. It is assumed that $h$ doesn't change with range, i.~e. the bottom is flat.
Using (\ref{BCs}), we disregard bottom attenuation directly.
However, sound absorption in the bottom is implicitly taken into account by means of a proper truncation of modal spectrum.
Particularly, we drop out all the modes which don't satisfy the condition
\begin{equation}
 E_m \le U(z=h),
\end{equation}
i.~e. only modes propagating without contact with the bottom are taken into account.

In the Hegewisch-Tomsovic method \cite{Tomc11,Hege13}, the propagator $\mathbf{G}(0,r_{\mathrm{f}})$ is expressed
as a product of propagators for intermediate segments of a waveguide:
\begin{equation}
\mathbf{G}(0,r_{\mathrm{f}}=Kr_{\text{b}})=\prod\limits_{k=0}^{K-1}\mathbf{G}_{K-k}((k-1)r_{\text{b}},kr_{\text{b}}).
 \label{BB}
\end{equation}
If the step $r_{\text{b}}$ is sufficiently large, segment propagators $\mathbf{G}_k$ with different $k$ are statistically independent from each other.
Furthermore, as the background sound-speed profile doesn't depend on range, one can assume that statistical properties of $\mathbf{G}$
are stationary along the waveguide. It yields $\mathbf{G}((k-1)r_{\text{b}},kr_{\text{b}}) = \mathbf{G}(r_{\text{b}})$.

A propagator for 
each individual segment can be calculated within the first-order perturbation theory, with the Cayley transform imposed to ensure unitarity.
The resulting formula is
\begin{equation}
\mathbf{G}(r_{\text{b}}) = \mathbf{\Lambda}[\mathbf{I} + i\mathbf{A}(r_{\text{b}})/2]^{-1}[\mathbf{I} - i\mathbf{A}(r_{\text{b}})/2].
\label{Cayley} 
\end{equation}
Here $\mathbf{I}$ is the identity matrix, and $\mathbf{\Lambda}$ is a diagonal matrix with elements
\begin{equation} 
 \Lambda_{mn} = \delta_{mn}e^{-ik_0E_mr_{\text{b}}},
 \end{equation} 
where $\delta_{mn}$ is the Kronecker symbol.
$\mathbf{A}$ is an inhomogeneity-induced perturbation matrix whose elements are calculated as
 \begin{equation}
  A_{mn}=k_0 \int\limits_{r'=0}^{r_{\text{b}}} e^{ik_0(E_m-E_n)r'}  V_{mn}(r')\,dr',
 \label{pert}
\end{equation}
\begin{equation}
V_{mn}(r) = \int \psi_m^*(z)V(r,z)\psi_n(z)\,dz.
 \label{Vmn}
\end{equation}
The key idea of the random matrix approach is to treat matrix elements of the perturbation $\mathbf{A}$ as 
random quantities
\begin{equation}
 A_{mn}(r_{\text{b}},k_0) = \sigma_{mn}(r_{\text{b}},k_0)z_{mn}(k_0),
 \label{Amn}
\end{equation}
where $\sigma_{mn}$ is calculated from spectral properties of the random inhomogeneity,
and $z_{mn}$ is a complex-valued Gaussian random variable with the unit variance.
It is important to note that variances $\sigma_{mn}$ can be found analytically (the corresponding formula is given in \cite{Hege13}).
The propagator step $r_{\text{b}}$ should be large enough to ensure statistical independence of propagators for neighboring segments. 
The upper bound for $r_{\text{b}}$ is determined by the condition $|A_{mn}|\ll 1$, otherwise the first-order perturbation theory doesn't apply.

Mode amplitudes of a wavefield can be combined into the vector $\vec{a}$, $\vec{a} \equiv (a_1, a_2, \text{...}, a_M)^T$.
In accordance with (\ref{evolution}), range evolution of this vector is governed by the equation
\begin{equation}
 \vec{a}(r) = \mathbf{G}(r)\vec{a}(0).
 \label{amod}
\end{equation}
It means that a wavefield can be calculated by means of sequential multiplication of the vector of mode amplitudes by the propagator matrix.
This algorithm is extremely fast if  number of propagating modes is not very large. 

\section{Model of a waveguide}
\label{Model}

In the present work we consider an acoustic waveguide in the deep ocean,
with an unperturbed sound-speed profile described by the biexponential model \cite{Chaos}
\begin{equation}
c_{\text{unpert}}(z)=c_0\biggl[\biggr.1+\frac{b^2}{2}\left(
e^{-az}-\eta
\right)^2\biggl.\biggr].
\label{BEP-prof}
\end{equation}
where $c_0=1490$~m/s,
$\eta =0.6065$, $a=0.5$~km$^{-1}$, $b=0.557$.
The biexponential profile closely resembles the celebrated canonical Munk model.

\begin{figure}[!ht]
\centerline{
  \includegraphics[width=.73\textwidth]{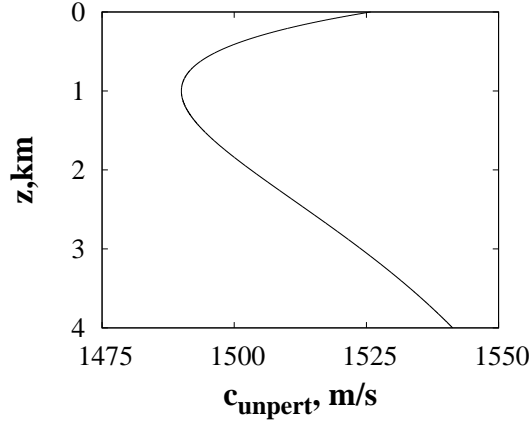}}
\caption{Biexponential sound-speed profile.
}
\label{Fig-BEP}
\end{figure}

We consider a large-scale inhomogeneity induced by a cold synoptic eddy.
The corresponding sound-speed perturbation is taken in the form \cite{Viro-Wamot,Viro-Akj08,Viro-Akj10,Radiophys}
\begin{equation}
\delta c_\text{lsc}=c_\mathrm{e}\exp\left(
-\frac{(r-r_{\mathrm{e}})^2}{2\Delta r^2}-\frac{(z-z_{\mathrm{e}})^2}{2\Delta z(r)^2}
\right),
\label{eddy}
\end{equation}
where
\begin{equation}
\Delta z(r)=\Delta z_{\text{c}}-\Delta z_{\upsilon}
\exp\left(-\frac{(r-r_{\upsilon})^2}{2\Delta r_{\upsilon}^2}\right).
\label{deltaz}
\end{equation}
The following parameter values are taken:
 $r_\mathrm{e}=250$~km, $z_{\mathrm{e}}=1$~km,
$\Delta r=120$~km, $\Delta z_{\mathrm{c}}=0.8$~km, $\Delta z_{\upsilon}=0.4$~km, 
$r_{\upsilon}=270$~km, $\Delta r_{\upsilon}=50$~km.

Sound-speed perturbation caused by internal waves is expressed as
\begin{equation}
 \delta c_{\text{iw}}(r,z) = c_0V_0\sum\limits_{j=1}^{j_{\max}} F_j(z) Y_{j}(r).
 \label{Vzr0}
\end{equation}
where $j_{\max}=50$,
\begin{equation}
 F_j(z) = \sqrt{\frac{1}{j^2+j_*^2}} e^{-3z/2B}\sin(j\pi\xi(z)),
 \label{Fj}
\end{equation}
$\xi(z) = e^{-z/B}-e^{-h/B}$, $B$ is the thermocline depth,
\begin{equation}
 Y_j(r) = \sum_l\sqrt{I_{j,l}}\cos(k_lr + \phi_{jl}),
 \label{Yi}
\end{equation}
$\phi_{jl}$ are random phases,
\begin{displaymath}
 V_0 = \frac{24.5}{g}\frac{2B}{\pi}N_0^2\sqrt{\frac{E\Delta k_l}{M}},
\end{displaymath}
$\Delta k_l$ is spacing between neighboring values of $k_l$.
This model was originally developed in \cite{ColBr98}.
Spectral weights $I_{j,l}$ are given by the formula
\begin{equation}
I(j,k_l) = \frac{k_j}{k_l^2+k_j^2} + \frac{1}{2}\frac{k_l^2}{(k_l^2+k_j^2)^{3/2}}
\text{ln}\frac{\sqrt{k_l^2+k_j^2}+k_j}{\sqrt{k_l^2+k_j^2}-k_j},
 \label{GM}
\end{equation}
where vertical wavenumbers are determined as
\begin{equation}
 k_j = \frac{\pi jf_{\text{i}}}{N_0B}.
\end{equation}
Formula (\ref{GM}) corresponds to the Garrett-Munk spectrum.
The following values of parameters are taken:
$N_0=2\pi/10$ min, $f_{\text{i}}=1$ cycle per day, the Garrett-Munk energy $E=6.3*10^{-5}$, mode scaling number $M=(\pi j_*-1)/2j_*^2$, 
and the principle mode number $j_*=3$.
We take 1000 values of
horizontal internal wave number $k_l$, which are equally spaced within the interval from $2\pi/100$ to $2\pi$ radians per km. 

Generally, vertical modes of internal waves depend on the horizontal wavenumber. In this case, the ansatz (\ref{Vzr0}) can be obtained by expanding
a random field $\delta c_{\text{iw}}$ over empirical orthogonal functions \cite{Radiophys}.

\section{The Hegewisch-Tomsovic method with adiabatic inhomogeneity imposed}
\label{Adiabatic}

Adiabatic variations of a waveguide can be taken into account in (\ref{Cayley}) by incorporating range dependence of normal modes and their eigenvalues.
Under some assumptions this range dependence can be evaluated using the perturbation theory \cite{LL3}, i. e. without solving the Sturm-Liouville problem
too frequently.
However, even in this way, statistics of integrals (\ref{Amn}) can be found only numerically, using Monte-Carlo sampling.
It remarkably increases computational time needed to estimate variances $\sigma_{mn}^2$. 
The situation becomes particularly worse if one uses the Hegewisch-Tomsovic method for modelling of acoustic pulses, 
when variances $\sigma_{mn}^2$ have to be computed for every frequency component. 

The problem can be partially resolved by optimizing the calculation of perturbation $V_{\text{iw}}$.
According to (\ref{Vzr0}), the function $V_{\text{iw}}$ is compound of many vertical modes, and amplitude of each vertical mode, $Y_j$, 
is commonly modelled as superposition of several hundred range harmonics. Calculation of $V_{\text{iw}}$ can be accelerated by representing
$Y_j$ as Fourier series 
\begin{equation}
 Y_j(r) = \sum_{n=-N}^N y_n^j e^{in\omega_{\text{b}} r}, \quad 
 \omega_{\text{b}} = \frac{2\pi}{r_{\text{b}}}.
\end{equation}
with random amplitudes $y_n$. Variance of $y_n$ can be estimated analytically:
\begin{equation}
 \sigma_y^2(j,n) = \frac{1}{4}\sum_l I_{j,l}\left[
 \text{sinc}^2\left(\frac{k_l-n\omega_b}{2}r_b\right) + \text{sinc}^2\left(\frac{k_l+n\omega_b}{2}r_b\right)
 \right].
 \label{sgmy}
\end{equation}
It turns out that number of Fourier harmonics needed for fair representation of amplitudes $Y_j(r)$
is about ten times smaller than number of harmonics in the expansion (\ref{Yi}).

Much more substantial reduction of computational cost is achieved by partitioning
a waveguide into short segments,
so that range variations of sound speed
due to the adiabatic term $V_{\text{lsc}}$ are negligible within each individual segment.
We can eliminate them by averaging:
\begin{equation}
 \bar U_k(z) = U(z) + \frac{1}{r_b}\int\limits_{(k-1)r_b}^{kr_b} V_{\text{lsc}}(r,z)\,dr.
\end{equation}
Then we can calculate local modes $\psi^{(k)}_m$ and eigenvalues $E^{(k)}_m$ by solving the Sturm-Liouville problem (\ref{StL})
with the averaged sound-speed profile $\bar U_k(z)$.
Variances $(\sigma_{mn}^{(k)})^2$ corresponding to the $k$-th segment now can be evaluated analytically:
\begin{equation}
 (\sigma_{mn}^{(k)})^2 = k_0^2r_b^2V_0^2\sum_j |F_{mn}^{jk}|^2\sum_{l=-L}^L\sigma_{y}^2(j,l)\text{sinc}^2\chi_{lmn}^{(k)},
 \label{varmn}
\end{equation}
where
\begin{equation}
 F_{mn}^{jk} = \int \psi_m^{(k)*}(z)F_j(z)\psi_n^{(k)}(z)\,dz, 
\end{equation}
\begin{displaymath}
 \chi_{lmn}^{(k)}\equiv\frac{(\omega_{mn}^{(k)}+l\omega_b)r_b}{2},\quad 
 \omega_{mn}^{(k)}\equiv k_0(E_m^{(k)} - E_n^{(k)}).
\end{displaymath}
Now we have to properly rewrite the formulae for the propagator construction from the preceding section:
\begin{equation}
 A_{mn}^{(k)}(r_{\text{b}}) = \sigma_{mn}^{(k)}(r_{\text{b}},k_0)z_{mn}^{(k)},
 \label{Amn2}
\end{equation}
\begin{equation}
\mathbf{G_k}(r_{\text{b}}) = \mathbf{\Lambda_k}[\mathbf{I} + i\mathbf{A_k}(r_{\text{b}})/2]^{-1}[\mathbf{I} - i\mathbf{A_k}(r_{\text{b}})/2],
\label{Cayley2} 
\end{equation}
where $\mathbf{A_k}$ is a random matrix consisted of elements $A_{mn}^{(k)}$, and $\mathbf{\Lambda_k}$ is a matrix with elements
\begin{equation} 
 \Lambda_{mn}^{(k)} = \delta_{mn}e^{-ik_0E_m^{(k)}r_{\text{b}}},
 \end{equation} 
Propagators $\mathbf{G_k}$ with different $k$ correspond to different basis sets of normal modes. 
As long as multiplication of two neighboring propagators requires them to be in the same basis,
the formula for the resulting propagator has to include an unitary matrix $\mathbf{S_k}$ for the basis transformation.
Elements of the transformation matrix are given by
\begin{equation}
 S_{mn}^{(k)} = \int \psi_m^{(k-1)}\psi_n^{(k)*}\,dz.
\end{equation}
Here it is assumed that the initial condition is taken as superposition of modes of an unperturbed waveguide,
and the matrix $\mathbf{S_1}$ describes basis transformation between unperturbed modes to modes of the first segment.
The resulting propagator reads
\begin{equation}
\mathbf{G}(Kr_b)=(\mathbf{G_K}\mathbf{S_K^{-1}})(\mathbf{G_{K-1}}\mathbf{S_{K-1}^{-1}})\text{...}(\mathbf{G_2}\mathbf{S_{2}^{-1}})(\mathbf{G_1}\mathbf {S_1})
\prod\limits_{k=1}^{K}\mathbf{S_{k}}.
\label{prop}
 \end{equation}
This equation corresponds to stepwise transformation of basis sets with increasing $k$.

\section{Numerical simulation}
\label{Numer}

\subsection{Intensity profile of a wavefield}
\label{Intensity}

\begin{figure}[!ht]
\begin{center}
  \includegraphics[width=.73\textwidth]{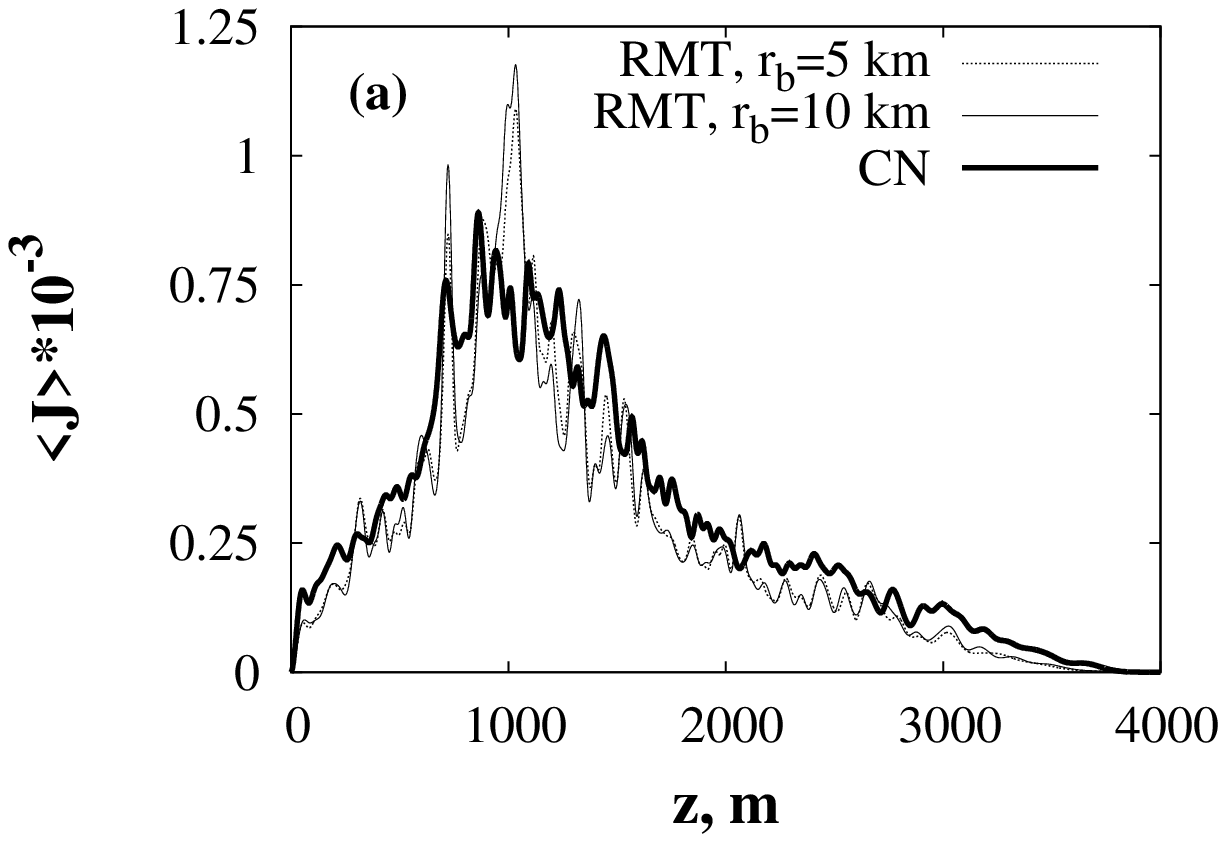}\\
  \includegraphics[width=.73\textwidth]{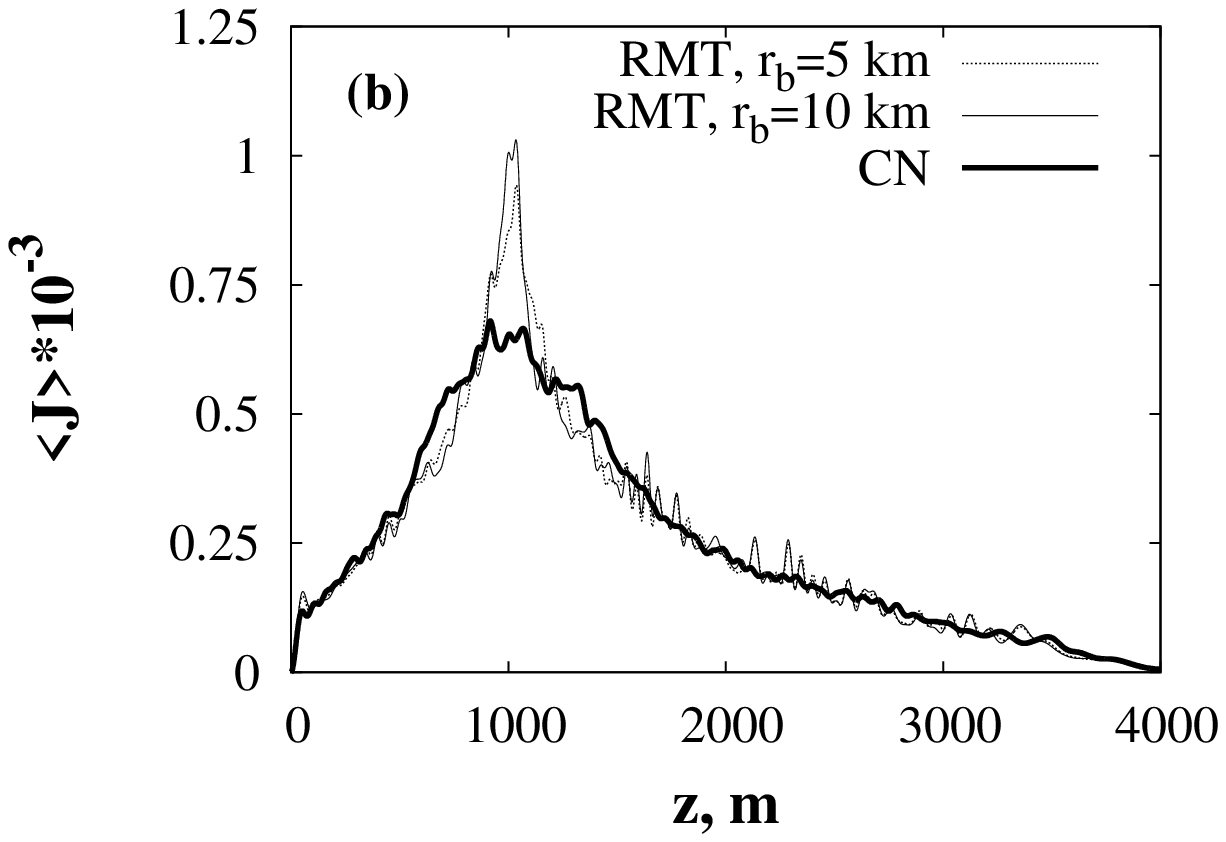}  
\end{center}  
\caption{Mean intensity of an acoustic wavefield 
as function of depth. (a) r = 200 km, (b) r = 500 km.
The curves obtained via the modified Hegewisch-Tomsovic method are denoted by ``RMT'', the curves denoted
``CN'' correspond to direct solution of the standard parabolic equation using the Crank-Nicholson scheme. Sound frequency is of 75 Hz.
}
\label{Fig2}
\end{figure}

Formula (\ref{prop}) was checked by means of numerical simulation with the model of a waveguide described in the Section \ref{Model}. 
Sound frequency was taken of 75 Hz. 
Computations were conducted for a point sound source located at the channel axis, $z=1$~km.
Figure \ref{Fig2} demonstrates the depth dependence of intensity, $J = |\Psi|^2$, averaged over 1000 realizations of an internal-wave field.
Direct solutions of the parabolic equation were obtained using the Crank-Nicholson scheme. 
It turns out that agreement between the modified Hegewisch-Tomsovic method and the Crank-Nicholson solutions improves
with increasing range. Probably, higher discrepancies for short ranges are  related to the presence of long-lasting horizontal correlations
that are ignored in the random matrix modelling.
Indeed, random matrix modelling implies that propagators for neighboring segments are statistically independent.
For $r=200$~km, the intensity profile corresponding to direct solution is significantly smoother than predictions
of the random matrix theory. 
In the case of $r=500$~km the difference is not so apparent, but one should notice that the modified
Hegewisch-Tomsovic method overestimates localization of a wavefield near the channel axis. Apart from the channel axis, the intensity profiles
almost coincide. Notably, the curve corresponding to random matrix modelling with $r_{\text{b}}=5$~km is smoother and closer to the curve
corresponding to direct solving than the curve corresponding to $r_{\text{b}}=10$~km. 
As long as reduction of $r_{\text{b}}$ makes the stepwise approximation of the propagator more accurate,
one may conclude that the presence of intensity oscillations imposed onto the smooth profile is associated with 
errors of the stepwise approximation.
In general, we see that the modified Hegewisch-Tomsovic method provides satisfactory agreement with direct solutions.

\subsection{Spectral statistics test}
\label{Spectral}

When we utilize any approximation, it is very important to ensure that it doesn't alter the underlying physics.
Information about physics of scattering is stored in spectrum of a wavefield propagator. 
It becomes evident if one invokes analogy with quantum mechanics, where spectral properties play a key role for dynamics.

We can check whether the modified Hegewisch-Tomsovic propagator (\ref{prop}) is able to reproduce
spectral statistics of the ``actual'' propagator obtained via the Crank-Nicholson scheme, or not.
Analysis of \cite{JCA} shows that spectral correspondence should be considered as a very stringent test, allowing one
to find out hidden discrepancies.

Eigenvalues and eigenfunctions of the propagator obey the equation
\begin{equation}
\hat G(0,r_F)\Phi_n(z)=g_n(r_0,r_F)\Phi_n(z).
\label{eigen}
\end{equation}
Owing to the unitarity of the propagator, eigenvalues can be recast as
\begin{equation}
 g_n=e^{-i\varphi_n}, \quad 
\varphi_n\in\Re.
\label{fn}
\end{equation}
This property means that the propagator matrix belongs to the so-called circular ensemble of random matrices \cite{Stockman}.
Scattering on random inhomogeneity reveals itself in statistics of level spacings \cite{UFN,PRE87}
\begin{equation}
\begin{gathered}
 s=\frac{k_0M(\varphi_{m+1}-\varphi_m)}{2\pi},\quad 
m = 1,2,\dots,M, \\
\varphi_{M+1} = \varphi_1 + \frac{2\pi}{k_0}.
\end{gathered}
\label{spacing}
\end{equation}
where the sequence of eigenphases $\varphi_m$ is rearranged  in the ascending order,
$M$ is the total number of eigenvalues for a single realization of the propagator, equal to the number
of propagating modes.
Statistical distribution of level spacings is connected to all $m$-order correlation functions of eigenvalues \cite{Stockman}.
Hence level spacing statistics serves as a good indicator of differences between the spectrum of the propagator constructed via random matrices
and the actual propagator obtained via the Crank-Nicholson scheme.

If scattering on inhomogeneity is weak, then the corresponding eigenphases of the propagator are statistically 
independent from each other, and level spacing distribution obeys the Poisson law
\begin{equation}
  \rho(s)\sim\exp(-s).
\label{Poisson}
\end{equation}
In the opposite case of strong scattering and global inter-mode coupling, the neighboring eigenphases ``repulse'' from each other \cite{Stockman}. It leads 
to level spacing statistics described by the Wigner surmise
\begin{equation}
\rho(s)\sim s^{\alpha}\exp\left(-Cs^2\right),
\label{surmise}
\end{equation}
where constants $\alpha$ and $C$ depend on symmetries of the propagator.
As the unitarity is the only constraint on the propagator, 
the propagator corresponds to the circular unitary ensemble (CUE). In this case we have
$\alpha=2$ and $C = 4/\pi$ \cite{Kol97}.

In the intermediate regime of moderate scattering one can use the Berry-Robnik distribution \cite{BR}
\begin{equation}
\rho(s)= \left[
v_{\mathrm{r}}^2\operatorname{erfc}\left(\frac{\sqrt{\pi}}{2}v_{\mathrm{c}}s\right)+\left(2v_{\mathrm{r}}v_{\mathrm{c}}+\frac{\pi}{2}v_{\mathrm{c}}^3s\right)
\exp\left(-\frac{\pi}{4}v_{\mathrm{c}}^2s^2\right)
\vphantom{\frac{\sqrt{\pi}}{2}}\right]\exp(-v_{\mathrm{r}}s),
\label{berrob}
\end{equation}
where  $v_{\mathrm{r}}+v_{\mathrm{c}}=1$.
Generally speaking, the Berry-Robnik formula (\ref{berrob}) is obtained under the assumption that the matrix $\mathbf{G}$
consists of two uncoupled blocks.
The first block is near-diagonal. It corresponds to weak scattering and regularly propagating modes.
The second block is a widely banded matrix, corresponding to strong scattering and ``chaotic'' modes.
Let's denote number of rows (or columns) in the first block as $M_{\text{r}}$. Then the parameters 
$v_{\mathrm{r}}$ and $v_{\mathrm{c}}$ are determined as
\begin{equation}
 v_{\mathrm{r}} = \frac{M_{\text{r}}}{M},\quad v_{\mathrm{c}}=\frac{M - M_{\text{r}}}{M} = \frac{M_{\text{c}}}{M}.
\end{equation}
Hence they can be thought of as fractions of weakly and strongly scattered modes, respectively.
Berry-Robnik distribution undergoes smooth transition
from the Poisson to the Wigner law as $v_{\mathrm{r}}$ decreases from 1 to 0.
Thus, fitting level spacing distribution by means of the formula (\ref{berrob}) and finding a value of $v_{\mathrm{r}}$ (or $v_{\mathrm{c}}$) corresponding
to the best fit, we can track the process of mode decoherence
due to scattering on random inhomogeneity.

\begin{figure}[!ht]
\begin{center}
  \includegraphics[width=.73\textwidth]{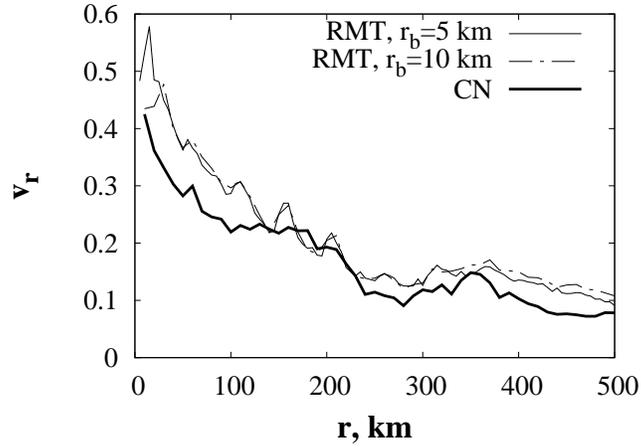}  
\end{center}  
\caption{Fraction of weakly scattered modes estimated using the Berry-Robnik formula (\ref{berrob}) vs range.
}
\label{Fig3}
\end{figure}

Figure \ref{Fig3} shows range dependence of the parameter $v_{\mathrm{r}}$. Apparently, the curves obtained via the modified Hegewisch-Tomsovic method
lie closely to the curve obtained via the Crank-Nicholson scheme.
However, we can see that the curve of the actual propagator corresponds to smaller values of $v_{\mathrm{r}}$ than predictions of 
the random matrix theory. It means that the latter ones slightly underestimate scattering.

The most intriguing feature of the curves presented in Fig.~\ref{Fig3} is increasing of $v_{\mathrm{r}}$ after crossing the synoptic eddy ($r\simeq 250$~km).
As $v_{\mathrm{r}}$ can be regarded as fraction of weakly scattered modes, it turns out that the eddy suppresses sound scattering.
Notably, this effect is well reproduced by the random matrix modelling.

\begin{figure}[!ht]
\begin{center}
  \includegraphics[width=.73\textwidth]{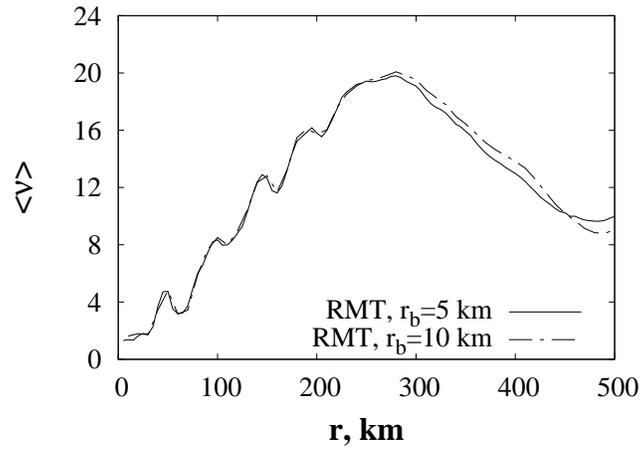}  
\end{center}  
\caption{Range dependence of mean participation ratio per eigenfunction of the propagator.
}
\label{Fig4}
\end{figure}

Strictly speaking, level spacing statistics cannot be considered as a absolutely reliable method of estimating scattering.
As it was shown in \cite{UFN}, the transformation of level spacing statistics back to the Poissonian form
may be caused by scattering on fine-scale structures and doesn't unambiguously indicate suppression of scattering. 
Therefore, identification of 
the mechanism responsible for such transformation
requires one to accompany the eigenvalue analysis by the analysis of propagator eigenfunctions.

Each eigenfunction can be expressed as superposition
of normal modes,
\begin{equation}
 \Phi_m(z) = \sum\limits_n b_{mn}\phi_n(z),
\label{eigf}
\end{equation}
where $b_{mn}$ is the $m$-th element of $n$-th eigenvector
of the matrix $\mathbf{G}$.
Scattering on random inhomogeneity leads to intense mode coupling. 
Consequently, a propagator eigenfunction corresponding to strong scattering should be compound of many normal modes.
Thus, we can estimate impact of scattering by exploring statistics of participation ratio values in the expansions (\ref{eigf}).
Participation ratio of the $n$-th eigenfunction is calculated as
\begin{equation}
\nu(n) = \left(
\sum\limits_{m=1}^M\lvert b_{mn}\rvert^4
\right)^{-1}.
 \label{npc}
\end{equation}
According to this definition,
$\nu$ is equal to 1 in a range-independent waveguide, and increases
as scattering intensifies.
Figure \ref{Fig4} demonstrates range dependence of participation ratio averaged over 
all eigenfunctions and realizations of random inhomogeneity.
After rapid growth for $r< 300$~km, mean participation ratio suddenly starts to decrease.
Hence the eigenfunction statistics 
confirms that growth of $v_{\mathrm{r}}$ is associated with suppression of scattering. 
These results anticipate a kind of anti-diffusive behavior, when some limited group of modes becomes more favorable for concentration of acoustic energy .

%
%

\section{Discussion}

Generalization of the Hegewisch-Tomsovic method onto waveguides involving large-scale inhomogeneity drastically extends range of its applications.
Indeed, real-world underwater acoustic waveguides are often subjected to longitudinal variations which can be treated as adiabatic.
It should be noted that efficiency of the method can be enhanced by using non-uniform partition of a waveguide.
Adjusting the propagator step with the rate of mesoscale variability, we can reduce inaccuracy of the stepwise approximation.
Furthermore, the modified Hegewisch-Tomsovic method looks as a promising tool for modelling in the presence of uncertainty in hydrological characteristics.

Nevertheless, the Hegewisch-Tomsovic method still has some limitations of the applicability. Firstly, the method is based on the perturbation theory and
can fail if it doesn't apply. It is the case, for example, for relatively high frequencies.
Secondly, the method relies on the narrow-angle approximation, therefore, it should not correctly incorporate wide-angle effects.
In this way it is reasonable to develop a version of the Hegewisch-Tomsovic method for the wide-angle parabolic equation, or for the Helmholtz equation. 
In the latter case formalism of S-matrices should be invoked \cite{Stockman}.

Correct calculation of matrix element variances is one of the main technical problems arising in the random matrix modelling.
Alternatively, these variances can be evaluated by solving the master equation for modal amplitudes \cite{DozierI,Creamer,ColosiMorozov,Colosi_Duda_Morozov}.
It is especially interesting in the context of the ``anti-diffusive'' behavior observed in this paper: 
can the master equation reproduce this effect?
It should be mentioned that
a somewhat similar behavior occurs in quantum systems,
when the so-called ``dark'' states accumulate population.
As it was shown in \cite{EPJB14,QE}, quantum master equation, being mathematically equivalent to the acoustical master equation,
readily reproduces this effect.
Therefore, it is reasonable to expect that the acoustical master equation can be reliable instrument for modeling 
using random matrices. 
It means that these two approaches can be efficiently combined.

\section{Conclusions}
\label{Concl}

The present paper is devoted to random matrix modelling of sound propagation in the ocean. It is shown that the approach of K.~Hegewisch and S.~Tomsovic
can be efficiently generalized onto waveguides with adiabatic inhomogeneity, even if magnitude of the inhomogeneity is relatively large. 
The generalization is obtained by means of stepwise approximation of the wavefield propagator, leading to the formula (\ref{prop}).
Efficiency of the modified Hegewisch-Tomsovic method is confirmed by numerical simulation for a model of an underwater sound channel
with a cold synoptic eddy imposed. Spectral analysis of the propagator has shown that the eddy leads to suppression of scattering on internal waves.

\section*{Acknowledgments}

This work was supported by the Russian Foundation of Basic Research within the projects 16-35-60040 and 16-05-01074, and  by 
the POI FEB RAS Program
'Mathematical simulation and analysis of dynamical processes in the ocean'
(№117030110034-7).
Author is grateful to Steven Tomsovic for stimulating and fruitful discussions.



\end{document}